\documentclass[prl,final,twocolumn,superscriptaddress]{revtex4}
\usepackage{graphicx}

\def\kcl{$\kappa$-Cl$\,\,$}

\newcommand{\mr}[1]{{{\mathrm{#1}}}}

\begin{document}

\title{Mott transition and transport crossovers
in the organic compound $\kappa$-(BEDT-TTF)$_{2}$Cu[N(CN)$_{2}$]Cl}

\author{P. Limelette}
\affiliation{Laboratoire de Physique des Solides (CNRS, U.R.A. 8502),
B\^atiment 510, Universit\'e de Paris-Sud, 91405 Orsay, France}
\author{P. Wzietek}
\affiliation{Laboratoire de Physique des Solides (CNRS, U.R.A. 8502),
B\^atiment 510, Universit\'e de Paris-Sud, 91405 Orsay, France}
\author{S. Florens}
\affiliation{LPT-Ecole Normale Sup\'erieure, (CNRS-UMR 8549),
24, rue Lhomond, 75231 Paris Cedex 05, France}
\affiliation{Laboratoire de Physique des Solides (CNRS, U.R.A. 8502),
B\^atiment 510, Universit\'e de Paris-Sud, 91405 Orsay, France}
\author{A. Georges}
\affiliation{LPT-Ecole Normale Sup\'erieure, (CNRS-UMR 8549),
24, rue Lhomond, 75231 Paris Cedex 05, France}
\affiliation{Laboratoire de Physique des Solides (CNRS, U.R.A. 8502),
B\^atiment 510, Universit\'e de Paris-Sud, 91405 Orsay, France}
\author{T.A.~Costi}
\affiliation{Institut f\"{u}r Theorie der Kondensierten Materie,
Universit\"{a}t Karlsruhe, 76128 Karlsruhe, Germany}
\author{C. Pasquier}
\affiliation{Laboratoire de Physique des Solides (CNRS, U.R.A. 8502),
B\^atiment 510, Universit\'e de Paris-Sud, 91405 Orsay, France}
\author{D. Jerome}
\affiliation{Laboratoire de Physique des Solides (CNRS, U.R.A. 8502),
B\^atiment 510, Universit\'e de Paris-Sud, 91405 Orsay, France}
\author{C. M\'ezi\`ere}
\affiliation{Laboratoire Chimie Inorganique, Mat\'eriaux et Interfaces (CIMI)
, FRE 2447 CNRS- Universit\'e d'Angers, 49045 Angers, France}
\author{P. Batail}
\affiliation{Laboratoire Chimie Inorganique, Mat\'eriaux et Interfaces (CIMI)
, FRE 2447 CNRS- Universit\'e d'Angers, 49045 Angers, France}

\begin{abstract}
\vspace{0.3cm}
We have performed in-plane transport measurements on the two-dimensional organic
salt $\kappa$-(BEDT-TTF)$_{2}$Cu[N(CN)$_{2}$]Cl. A variable (gas) pressure
technique allows for a detailed study of the changes in conductivity through the
insulator-to-metal transition. We identify four different transport regimes as a
function of pressure and temperature (corresponding to insulating, semi-conducting,
''bad metal'', and strongly correlated Fermi liquid behaviours).
Marked hysteresis is found in the transition region, which displays complex
physics that we attribute to strong spatial inhomogeneities.
Away from the critical region, good agreement is found with a dynamical
mean-field calculation of transport properties using the numerical renormalization
group technique.
\end{abstract}

\maketitle

The Mott metal-insulator transition (MIT) is a key phenomenon in the
physics of strongly correlated electron materials.
It has been the subject of extensive experimental studies in transition metal oxides, such as
(V$_{1-x}$Cr$_x$)$_2$O$_{3}$
or chalcogenides such as NiS$_{2-x}$Se$_x$ (for a review, see Ref.~\cite{imada}).
In contrast to chemical composition, hydrostatic pressure allows in principle to
sweep continuously through the transition.
For these materials however, the appropriate range of pressure is several kilobars.
For this reason, many fundamental aspects of the MIT are yet to be studied
in detail. This issue is particularly important in view of
recent theoretical predictions for e.g. spectroscopy and transport close to the
transition, which should be put to experimental test \cite{Georges96}.
Layered charge-transfer salts of the $\kappa$-(BEDT-TTF)$_{2}\mr{X}$ family
(where X is a monoanion)
offer a remarkable opportunity for such a study. Indeed, these compounds
are known to display a great sensitivity to hydrostatic
pressure \cite{Ito96}. The $\kappa$-(BEDT-TTF)$_{2}$Cu[N(CN)$_{2}$]Cl compound
in particular (abbreviated \kcl below) displays a very rich phase diagram
with paramagnetic insulating, antiferromagnetic insulating, superconducting
and metallic phases when pressure is varied over a range of a few hundred
bars \cite{McKenzie97,Lef00,Ito96}.

\begin{figure}[htbp]
\centerline{\includegraphics[width=0.85\hsize]{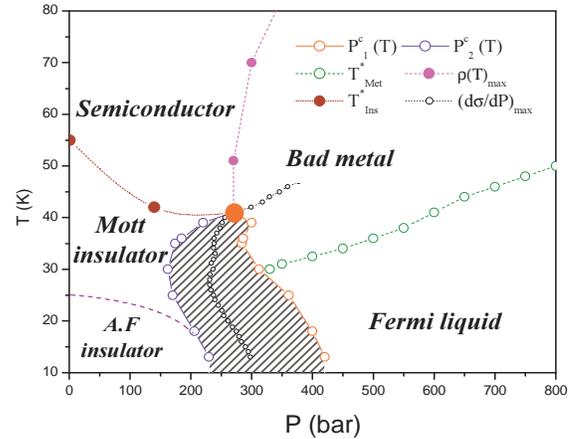}}
\caption{Pressure-Temperature phase diagram of the \kcl salt.
The crossover lines identified from our transport measurements
delimit four regions, as described in the text.
The spinodal lines defining the region of coexistence
of insulating and metallic phases (hatched) are indicated, as well as
the line where $\mr{d}\sigma/\mr{d}P$ is maximum. The latter yields an
estimate of the first-order transition line, ending in a critical endpoint.
The transition line into an antiferromagnetic insulating phase has been taken
from Ref.~\cite{Lef00}, while the superconducting phase \cite{Ito96,Lef00}
below $13$~K has been omitted.}
\label{fig-ph-diag}
\end{figure}

\begin{figure}[htbp]
\centerline{\includegraphics[width=0.95\hsize]{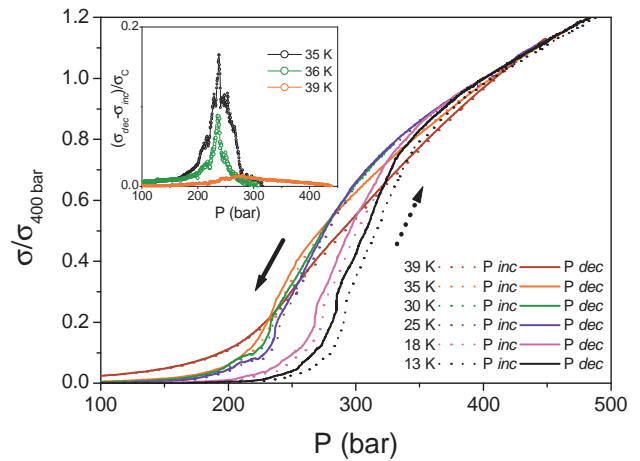}}
\caption{Conductivity measured for increasing ($\sigma_{inc}$) and
decreasing ($\sigma_{dec}$) pressure sweeps,
at different temperatures up to 39K. Hysteresis at 35 K and
39 K is more apparent in the difference $\sigma_{dec}-\sigma_{inc}$ (inset).}
\label{fig-hyst}
\end{figure}

In this work, we report on an extensive experimental study of the in-plane
resistivity of the \kcl compound for a range of pressure spanning both
the insulating and the metallic phases.
In contrast to previous studies \cite{Ito96}, pressure is varied continuously
using a Helium gas cell, at constant temperature.
By analyzing the pressure and temperature dependence of the measured resistivity,
we have identified important crossover lines, which are
summarized on the phase diagram of Fig.\ref{fig-ph-diag}. These crossovers separate
four different regimes of transport (to be described below) within the paramagnetic phase,
corresponding to an ''insulator'', a ''semiconductor'', a ''bad metal'' and a Fermi-liquid
metallic regime.
Our measurements were performed both with increasing and decreasing
pressure sweeps, yielding a determination of the two spinodal
lines $P_1^c(T)$ and $P_2^c(T)$ shown in Fig.~\ref{fig-ph-diag}
(see also Fig.~\ref{fig-hyst}).
%
A critical comparison is made to dynamical mean-field theory (DMFT \cite{Georges96}),
which describes successfully the observed transport crossovers, while
the critical region itself reveals more complex behaviour.
Numerical Renormalization Group (NRG) calculations of the resistivity 
within DMFT are reported here for the first time, and compare
favorably to the experimental data.

\begin{figure}[htbp]
\centerline{\includegraphics[width=0.95\hsize]{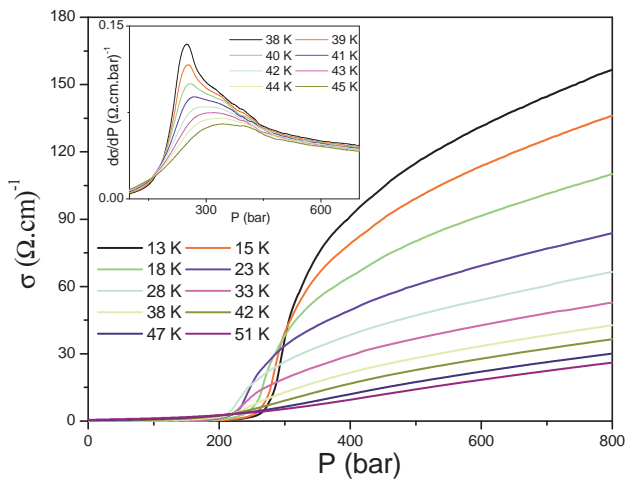}}
\caption{Pressure-dependence of the in-plane conductivity $\sigma_{ab}$ at different temperatures.
The inset shows the derivative $\mr{d}\sigma_{ab}/\mr{dP}$ which reveals a sharp peak at low
temperature.}
\label{fig-rhovsp}
\end{figure}

This paper is based mainly on the data set displayed on
Fig.~\ref{fig-rhovsp}. The in-plane conductivity was measured using a standard
four- terminal method as a function of
pressure between 1 bar and 1 kilobar. The temperature range 13K-52K was investigated, with
pressure sweeps performed every Kelvin (for clarity, only selected
temperatures are displayed on Fig.~\ref{fig-rhovsp}).
The data clearly demonstrate how pressure drives the system from a low-conductivity
insulating regime at low pressure (below 200 bar) to a high-conductivity metallic regime at
high pressure (above 350 bar), with a more complex transition region
in the 200-350~bar range.

The pressure interval in which the normalized difference between these two
measurements exceeds our experimental precision
(inset of Fig.~\ref{fig-hyst}) corresponds to the region of coexistence between the
insulating and metallic phases. This interval is found to be rather large
($\sim 100$~bar) at $35$~K, and still measurable (although the signal is
close to our error bars) at $39$~K. Note that a much narrower interval
would be found from the width at half-maximum.
We emphasize that varying pressure rather than temperature is a much more accurate
experimental technique for the investigation of the coexistence region, because of the
shape of the spinodal lines in the $(P,T)$ plane. This also ensures
a better accuracy on pressure (of order 1 bar) as compared to a cooling experiment.
Our measurements yield a determination of the critical point in the
$T_c\simeq 39$~K to $T_c\simeq 42$~K range
($P_c\simeq 280$~bar), somewhat higher than previously
reported from NMR \cite{Lef00} or sound velocity  \cite{Fou02} experiments.
The data in Fig.~\ref{fig-hyst} are far from the ''ideal'' situation of
a sharp discontinuity in $\sigma(p)$ for $T<T_c$,
and present rather a wide region of phase coexistence, in which the measured $\sigma(P)$
displays sizeable noise.
This suggests a complex physics in this regime, with the likely
appearance of insulating and metallic domains in the material, and pinning by
disorder.

\begin{figure}[htbp]
\centerline{\includegraphics[width=0.95\hsize]{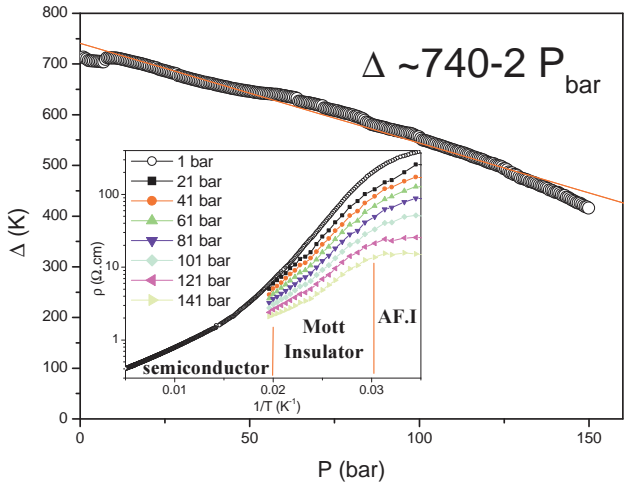}}
\caption{Pressure-dependence of the paramagnetic gap, as determined from the activation
plot shown in the inset.}
\label{fig-gap}
\end{figure}

We now focus on the low-pressure,
insulating regime. In the inset of Fig.~\ref{fig-gap}, the conductivity data in this
pressure range have been replotted as $\log\rho$ {\it vs} $1/T$.
At lower temperature (below 50 K), an activation law
$\rho\sim \exp(\Delta/2T)$ describes the data quite well.
The temperature range over which this activation law applies is limited both
from below and from above. Indeed, at the lowest temperatures a transition into the
antiferromagnetic insulating phase is found (around 25 K at ambient pressure, see
Fig.~\ref{fig-ph-diag}), while a crossover to a different insulating regime is
observed at higher temperatures, as discussed below.
Despite this limited range, a reasonable determination of the insulating gap
$\Delta$ in the ''paramagnetic insulator'' region of Fig.~\ref{fig-ph-diag}
can be achieved, since $\Delta\gg T$.
An approximately linear pressure dependence (see Fig.~\ref{fig-gap}) is found:
$\Delta\simeq 740-2 P_\mr{bar}$ (in Kelvin).
Hence, the activation gap is still {\it a large scale} close to the coexistence
region ($\Delta \simeq 400\,\mbox{K}$ at $150$~bar), of the order of ten times the critical temperature
associated with the transition. $\Delta$ cannot be reliably determined above $150$~bar,
but a rough extrapolation
would lead to a gap closure around $370$~bar, on the {\it high-pressure side} of the coexistence
region of Fig.~\ref{fig-ph-diag}, suggesting
that the finite-temperature Mott transition in this material is {\it not driven}
by the closure of the Mott gap.
At higher temperature (above $T\simeq 50$~K at ambiant pressure), the $T$-dependence
of the resistivity deviates from the above activation law (inset of Fig.~\ref{fig-gap}).
This crossover into a ''semi-conducting'' regime, which extends over much of the high-T part of the phase
diagram, is indicated on Fig.~\ref{fig-ph-diag}.
Interestingly, the crossover scale is {\it an order of magnitude
smaller} than the insulating gap.

\begin{figure}[htbp]
\centerline{\includegraphics[width=0.95\hsize]{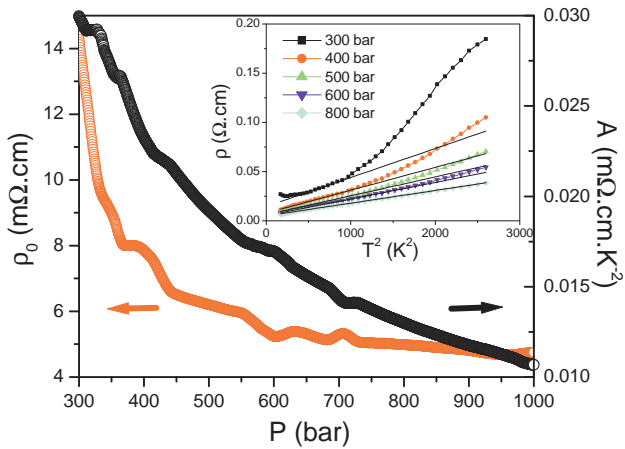}}
\caption{Pressure-dependence of the $T^2$ coefficient $A$ and residual
resistivity $\rho_0$.
Inset: Plot of $\rho$ {\it vs.} $T^2$, also showing the increase of $T^*$ with
pressure (straight lines are guides to the eyes).}
\label{fig-T2}
\end{figure}

The high pressure regime above $300$~bar is now considered. As demonstrated in
the right inset of Fig.~\ref{fig-T2}, the resistivity in this regime has a
quadratic, Fermi liquid dependence upon temperature at low temperatures:
$\rho \!=\! \rho_0(P) + A(P)\,T^2$. The quality of the fit, obtained by replotting the data
set of Fig.\ref{fig-rhovsp} as a function of $T^2$,
illustrates the high precision of the variable pressure technique.
The coherence temperature $T^*$  above which this law is no longer valid (of the order
of $35$~K at $500$~bar) defines a crossover into a
''bad metal'' regime, as indicated on Fig.~\ref{fig-ph-diag}.
The prefactor $A(P)$ of the $T^2$ dependence is found to depend
strongly on pressure, as displayed on Fig.~\ref{fig-T2}, and
the product $A(P)\times(T^*)^2$ remains approximately pressure-independent.
These findings correspond to a strongly correlated Fermi liquid regime at low
temperature.
We note that the residual resistivity has a weaker pressure-dependence than A,
but does increase close to the coexistence region.
While $A(P)$ cannot be determined precisely below $280$~bar, the data
suggest a divergency of $A$ at $P\simeq 220$~bar, significantly smaller than the pressure
at which the extrapolated insulating gap would vanish (of order $370$~bar, see above).
This suggests that the closure of the Mott-Hubbard gap and the loss of Fermi liquid
coherence are two distinct phenomena, associated with very different energy
scales, as is also clear from the fact that the coherence scale $T^*$
(a few tens of Kelvin) is much smaller than the insulating gap $\Delta$
(several hundreds Kelvin).

\begin{figure}[htbp]
\centerline{\includegraphics[width=0.95\hsize]{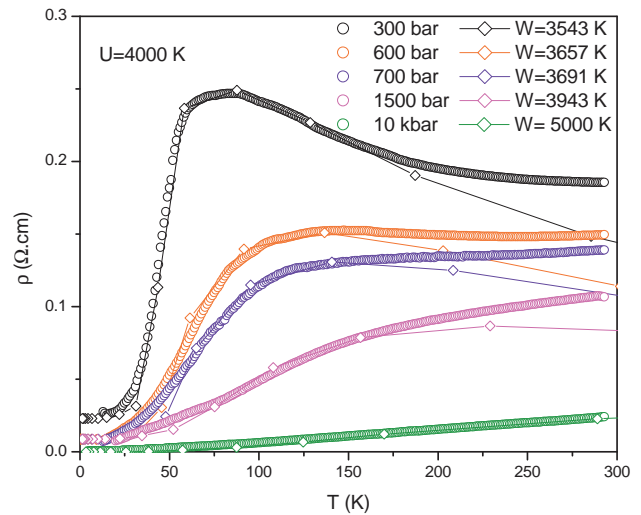}}
\caption{Temperature-dependence of the resistivity at different pressures.
The data (circles) are compared to a DMFT-NRG calculation (diamonds), with a
pressure dependence of the bandwidth as indicated. The measured residual
resistivity $\rho_0$ has been added to the theoretical curves.}
\label{fig-r(T)}
\end{figure}

In order to better characterize the crossover into the ''bad metal''
as temperature is increased above $T^*$,
we have performed measurements in a wider temperature range, up to $300$~K, as
displayed on Fig.~\ref{fig-r(T)}.
We confirm the general behaviour reported by other authors \cite{Ito96}. At moderate
pressures (a few hundred bars), the resistivity changes from a $T^2$ behaviour
below $T^*$, to a regime characterized by very large values of the resistivity
(exceeding the Ioffe-Regel-Mott limit by more than an order of magnitude)
but still metallic-like ($\mr{d}\rho/\mr{d}T\!>\!0$).
For pressures in the transition region, this increase persists
until a maximum is reached, beyond which a ''semi-conducting'' regime is entered
($\mr{d}\rho/\mr{d}T\!<\!0$).
This regime is continuously connected to that
found on the insulating side, at temperatures above $50$~K.
Fig.~\ref{fig-ph-diag} summarizes the four regimes of transport which we have described
so far.

We now compare these experimental findings to the DMFT description of the Mott transition.
Similarities between DMFT and some physical properties of BEDT organics have
been emphasized previously \cite{McKenzie98,Mer00}.
One of the key qualitative outcomes of our measurements is the {\it separation of
energy scales} found in the transition region. This is also a distinctive feature
of DMFT, in which Hubbard bands are well separated from the quasi-particle
peak in the correlated metal, and a coexistence region of insulating and metallic
solutions delimited by two spinodal lines is found.
The three transport regimes found on the
metallic side of the transition are well accounted for within DMFT
\cite{Roz94,Maj95,Georges96,Mer00}.
In this theory, a Fermi- liquid regime with long lived quasiparticles applies
below the coherence scale $T^*$, with $\rho=A T^2$ and
$A\propto 1/(T^*)^2$. As $T$ approaches $T^*$, the lifetime decreases sharply while
the low-energy local density of states (d.o.s) is strongly depleted. This leads to a
large resistivity, increasing rapidly with $T$, characteristic of a bad metal.
For $T\gg T^*$, the local d.o.s displays a pseudogap
and a ''semi-conducting'' transport regime is found, with $d\rho/dT<0$.
We have performed DMFT calculations of the resistivity
for a simple Hubbard model at half-filling. The NRG technique was used \cite{Bul01},
which is crucial to capture correctly the enhancement of $A$, and to
determine accurately the resistivity maximum.
The coupling was fixed at $U=4000$~K, consistent with the measured critical temperature
(within DMFT, $T_c\simeq U/100$). A semi-circular band was used,
with a bandwidth $W$ adjusted for each pressure, as indicated on
Fig.~\ref{fig-r(T)}. Comparison to the measured $\rho(T)-\rho_0$
involves also a global scale factor (the same for all curves).
We find a good agreement between our calculations and the
data, up to $T\sim 150$~K, for pressures ranging from the transition
(300 bar) to $10$~kbar. The fitted bandwidth increases by about $40\%$ in that range.
This value, together with the magnitudes of $U$ and $W$ is consistent with
theoretical \cite{Rahal97} and experimental \cite{Kle00} estimates for these materials.
We attribute the discrepancy observed above $\sim 150$~K to
the strong thermal dilatation of the materials, leading to a smaller effective
bandwidth, an effect which could be taken into account by allowing for
a T-dependence of $W$ in the calculation.
The same effect might also contribute to the observed crossover on the
insulating side at $T\sim 50$~K, even though DMFT also leads to a similar
prediction of a purely electronic crossover at $T\ll\Delta$ \cite{Roz94,Georges96}.

Finally, we observe that our data {\it in the critical regime} 
around $(T_c,P_c)$ do not appear to obey the
critical behaviour established within DMFT,
namely that of a liquid-gas transition in the
Ising universality class \cite{Kot00}. The observed transition is much too smooth, and the coexistence
region too wide, to be described in that manner.
Besides the low-dimensionality of these compounds, this
could be attributed to strong spatial inhomogeneities or disorder in the transition region.

To conclude, we have performed detailed transport measurements on the \kcl compound
using a variable pressure technique. Four transport regimes have been identified as a function
of temperature and pressure, separated by crossover lines. Hysteresis experiments have confirmed
the first-order nature of the transition and allowed for the first detailed determination of
the coexistence region. DMFT provides a good qualitative description of these crossovers,
and new DMFT-NRG calculations of transport are in good agreement with the data.
Further theoretical work is needed to understand the critical regime, however.
We note that recent ultrasound velocity measurements \cite{Fou02} have revealed anomalies
along crossover lines very
similar to the two high-temperature lines reported here from transport measurements.
This calls for further theoretical work aiming at connecting these two effects.

\begin{acknowledgments}
We are grateful to R. Chitra, V. Dobrosavlevic, G. Kotliar, R. McKenzie,
M. Poirier, M. Rozenberg and A.M. Tremblay for useful
discussions. T.A.C., S.F. and A.G. are grateful to KITP-UCSB for hospitality during the
final stages of this work (under NSF grant PHY99-07949). T.A.C. acknowledges
support from the SFB 195 of the DFG.
\end{acknowledgments}

\end{document}